\documentclass[lettersize,journal]{IEEEtran}
\usepackage{amsmath,amsfonts}
\usepackage{amsthm}
\usepackage{thmtools}
\usepackage{chngcntr}
\usepackage{colortbl}
\usepackage{algorithmic}
\usepackage{algorithm}
\usepackage{array}
\usepackage{textcomp}
\usepackage{stfloats}
\usepackage{url}
\usepackage{verbatim}
\usepackage{graphicx}
\usepackage{cite}
\usepackage{bm}
\usepackage[dvipsnames]{xcolor}
\usepackage{braket}
\usepackage{caption}
\usepackage{subcaption}
\usepackage[colorlinks=true,
            linkcolor=red,
            urlcolor=blue,
            citecolor=gray]{hyperref}

\definecolor{noattack}{rgb}{0.24,0.74,0.76}
\definecolor{Minor}{rgb}{0.0,0.62,0.42}
\definecolor{Moderate}{rgb}{0.96,0.84,0.34}
\definecolor{Moderate}{rgb}{1.0,0.75,0.0}
\definecolor{Moderate}{rgb}{1.0,0.84,0.07}
\definecolor{Moderate}{rgb}{1.0,0.76,0.07}
\definecolor{Severe}{rgb}{1.0,0.49,0.01}
\definecolor{Severe}{rgb}{1.0,0.53,0.01}
\definecolor{Critical}{rgb}{0.88,0.02,0.02}

\definecolor{noattackrow}{rgb}{1,1,1}
\definecolor{Minorrow}{rgb}{0.66,0.89,0.63}
\definecolor{Minorrow}{rgb}{0.67,0.94,0.82}
\definecolor{Minorrow}{rgb}{0.83,0.98,0.91}
\definecolor{Moderaterow}{rgb}{0.96,0.84,0.34}
\definecolor{Moderaterow}{rgb}{0.97,0.91,0.56}
\definecolor{Moderaterow}{rgb}{1,0.97,0.81}
\definecolor{Severerow}{rgb}{1.0,0.75,0.00}
\definecolor{Severerow}{rgb}{1.0,0.83,0.68}
\definecolor{Criticalrow}{rgb}{0.94,0.5,0.5}
\definecolor{Criticalrow}{rgb}{1.0,0.6,0.6}
\definecolor{Criticalrow}{rgb}{1.0,0.65,0.65}
\usepackage{titlesec}
\titlespacing*{\section}{0pt}{10pt}{5pt}   
\titlespacing*{\subsection}{0pt}{5pt}{5pt}  

\begin{document}

\title{SWAP Attack: Stealthy Side-Channel Attack on Multi-Tenant Quantum Cloud System}

\author{Wei Jie Bryan Lee$^{1}$, Siyi Wang$^{1}$, Suman Dutta$^{1,2}$, Walid El Maouaki$^{3}$, Anupam Chattopadhyay$^{1}$\\
        \small $^{1}$College of Computing and Data Science, Nanyang Technological University, Singapore \\
        \small $^{2}$Applied Statistics Unit, Indian Statistical Institute, Kolkata, India\\
        \small $^{3}$Quantum Physics and Spintronics Team, LPMC, Faculty of Sciences Ben M’Sik, Hassan II University of Casablanca, Morocco \\
}





\maketitle

\begin{abstract}
The rapid advancement of quantum computing has spurred widespread adoption, with cloud-based quantum devices gaining traction in academia and industry. This shift raises critical concerns about the privacy and security of computations on shared, multi-tenant quantum platforms accessed remotely.
In several recent works, crosstalk has been exploited on shared quantum devices, enabling adversaries to interfere with the victim circuits within a neighborhood. Though illuminating, these studies left several unresolved questions regarding the fundamental cause of crosstalk, effective countermeasures, and replicability across different circuits.
In this paper, we revisit the crosstalk effect, tracing its origins to the SWAP path between various qubits in a circuit and demonstrating its effectiveness even over long distances. The results significantly improve our understanding of the phenomenon beyond prior works.
Our proposed SWAP-based side-channel attack can be executed in both active and passive modes, as verified on real shared IBM quantum devices.
In the active attack, an adversary executing a single CNOT gate can perturb victim circuits running Grover's Algorithm from afar, reducing the expected output accuracy through $81.62\%$ by strategically positioning qubits.
Moreover, this phenomenon can be clearly explained through modeling, where specific qubits in a platform can be identified as more susceptible to an attacker.
The passive attack, leveraging a stealthy circuit as small as $6.25\%$ of the victim's, achieves $100\%$ accuracy in predicting the circuit size of the victim, running Simon's Algorithm.
These findings challenge the existing defense strategy of maximizing topological distance between circuits, demonstrating that attackers can still extract sensitive information or manipulate results remotely. Our work highlights the need for more robust security measures to safeguard quantum computations against such emerging threats.
\end{abstract}


\section{Introduction}~\label{Introduction}
Quantum computing has witnessed significant advancements in recent years~\cite{bravyi_code,quantum_supremacy,acharya2024quantum}, driven by hardware breakthroughs~\cite{acharya2024quantum,main_distributed_quantum}, new circuits for various applications~\cite{wang2025comprehensive,Gidney2021howtofactorbit,baksi_quantum_aes,constant_depth_ml_activation} and increased accessibility through cloud services. Indeed, major cloud providers such as IBM Quantum~\cite{ibm_quantum_platform, Santos_2016}, AWS Braket~\cite{amazon_braket, aws}, and Microsoft Azure Quantum~\cite{microsoft_azure_quantum} have democratized access to quantum hardware, enabling researchers and developers to experiment with real quantum processors. IBM continues to push the boundaries with its road map toward error-corrected quantum computing, while AWS Braket and Azure integrate various quantum backends, fostering innovation across academia and industry. With the growing adoption of cloud-based quantum computing, privacy, and security have gained greater prevalence and importance in our current day and age.

In this work, we propose a novel SWAP-based side-channel attack that can be executed in both active and passive modes. Our investigation reveals that the noise generated within quantum circuits originates from the SWAP path between qubits, which remains highly effective even over long distances.

In active SWAP attacks, indirect connectivity between topologically distant qubits enables attackers to perturb a victim's circuit without detection. This attack provides granular control, ranging from minor to critical disruption; rendering previously suggested defense strategies—such as maintaining a topological distance between circuits—ineffective. Consequently, victim circuits remain vulnerable even under existing security measures, posing a significant risk in real-world cloud-based quantum computing environments where multiple users share common resources.

In passive SWAP attacks, adversaries exploit side-channel leakage to extract sensitive information from an unsuspecting victim's circuit.
Due to improper qubit allocation in shared quantum devices, certain circuits become vulnerable to unintended information leakage. This attack leverages crosstalk noise to predict details such as the structure and size of a victim's circuit.

All proposed attacks have been experimentally validated on real cloud-based quantum devices, first using \textit{ibm\_kyoto}, and later migrated to \textit{ibm\_brisbane}, where the results align with our theoretical analysis.
%

In brief, this work makes the following contributions:
\begin{itemize}
\item \textbf{Active SWAP Attack} that can be executed from a distance, causing varying levels of disruption—from minor, hard-to-detect interference to critical attacks that significantly impair the victim's circuit.
\item \textbf{Passive SWAP Attack} that exploits improper qubit allocation, leading to leakage of sensitive information from the victim's circuit. Additionally, we analyze the trade-off between the attacker's circuit size and the prediction accuracy and confidence.
\item \textbf{Challenging Existing Defenses}: This novel distant active SWAP attack method challenges the existing defense strategy, which assumes that maintaining a topological distance between the victim and attacker circuits ensures security.
\end{itemize}
%

This paper is organized as follows: Section~\ref{Previous} reviews prior work on quantum crosstalk attacks. Subsequently, Section~\ref{AttackModel} details the attack model and Section~\ref{Methodolgy} then outlines our attack methodology, detailing the active and passive SWAP attacks. Section~\ref{Results} presents experimental findings 
and provides insights into the results. Finally, Section~\ref{Future} concludes the paper with potential research directions.

\section{Previous Works}~\label{Previous}

In recent years, quantum cybersecurity has seen rapid advancements, with extensive research exploring various security challenges~\cite{Quantum_Cloud_Attack1, xu2023exploration, bell2022reconstructing, wang2024poster}. Among these, the security of multi-tenant quantum cloud computing has emerged as a critical concern.

The concept of multi-tenant quantum computing, introduced to enhance resource utilization, presents new challenges to the security of quantum computations. One critical vulnerability in such environments arises from crosstalk, which adversaries can exploit to manipulate the output of a victim's circuit.

Quantum crosstalk refers to unintended interactions between qubits during parallel gate operations in a quantum processor, resulting from physical connectivity and coupling. These interactions allow operations on one set of qubits to influence nearby qubits, thereby reducing fidelity and introducing errors that degrade computational performance \cite{murali2020software}. Understanding and mitigating crosstalk is essential for the development of reliable and scalable quantum computers.

In 2020, Sarovar et al. \cite{sarovar2020detecting} introduced a comprehensive framework for identifying and characterizing crosstalk errors, providing a rigorous definition and a protocol for their detection and localization—crucial for mitigation strategies.

One of the earliest studies on noise-induced vulnerabilities in multi-tenant quantum computing environments is Saki et al. \cite{saki2021impact}, which provided a detailed analysis of noise types in quantum systems, including crosstalk and highlighted their role in security risks such as fault injection and data leakage. The authors proposed program isolation and qubit reallocation strategies to mitigate these threats.
Subsequently, Zhao et al. \cite{zhao22} analyzed quantum crosstalk in superconducting quantum processors with fixed-frequency transmon qubits coupled via a tunable bus, highlighting how residual inter-qubit coupling induces errors in simultaneous gate operations and emphasizing the need for optimized qubit architectures. Additionally, Winick et al. \cite{winick21} proposed a scalable framework for modeling crosstalk effects, demonstrating how optimal control techniques can enable high-fidelity parallel operations despite substantial local and non-local crosstalk. In 2023, Zhou et al. \cite{zhou23} developed an analytical condition for crosstalk-robust single-qubit control in multi-qubit systems using cumulant expansion to suppress leading-order crosstalk contributions, thereby improving multi-qubit characterization and control.
Furthermore, an extensive characterization of crosstalk noise in IBM quantum devices was conducted in Harper et al. \cite{harper2024crosstalk}, demonstrating how adversaries can exploit proximate attacker qubits to disrupt a victim's quantum circuit outputs via crosstalk noise. The study proposes defense strategies, including circuit separation, which minimizes crosstalk-induced errors by increasing the physical distance between sensitive and potentially disruptive qubits.


While earlier research has mostly concentrated on proximity-based attacks, our method challenges existing defense measures, particularly qubit separation techniques, by introducing a novel attack model based on non-local interactions. Unlike previous proximity-based approaches, which use neighboring qubits to allow more substantial and localized interactions, our strategy deliberately separates the attacker's qubits on the quantum hardware, ensuring they are not physically adjacent to the target qubits. This increases the complexity of the attack while also disrupting typical proximity-based optimization tactics.

Although we were unable to replicate previous works on the \textit{ibm\_brisbane} quantum device, the proposed SWAP-induced attacks would render previously proposed defense strategies, such as enforcing a topological distance between circuits, ineffective. Furthermore, for the active SWAP attack, we introduce a lightweight attack circuit that improves upon prior designs, which relied on multiple CNOT gates to intersect a potential SWAP path. In contrast, our approach requires only a single CNOT gate while still inflicting significant disruption on victim circuits.

\section{Attack Model}~\label{AttackModel}
In this section, we specify the attack objectives and outlines the assumptions regarding the attacker’s capabilities.  

\subsection{Attack Objective}
Although both attacks exploit the SWAP path vulnerability, the objectives of the active and passive attacks are different. The details are outlined as follows:
\begin{itemize}
    \item In the active SWAP attack, the attacker aims to introduce errors into the victim’s output by executing a CNOT gate concurrently with the victim’s two-qubit Grover’s circuit on a cloud-based multi-tenant quantum system.  
    \item In contrast, the objective of the passive SWAP attack is to extract sensitive information from the victim’s circuit. In our experimental configuration, the victim runs Simon’s algorithm with an unknown shift $s$. The attacker’s goal is to infer critical attributes of the hidden shift, such as its size and output.  
\end{itemize}


\subsection{Attack Capacity}
To execute both active and passive attacks, we assume that the attacker possesses the following capabilities:
\begin{itemize}
\item The attacker can run its circuit on a multi-tenant quantum device (e.g., \textit{ibm\_brisbane} with $127$ qubits) simultaneously with the victim’s circuit.
\item For all circuit transpilation processes, it is assumed that the SWAP paths remain consistent across all qubits in the circuit.
\item The attacker is assumed to conduct the attack within the same calibration window during which the noise patterns were generated. This ensures that any observed effects result from the intersecting SWAP path, rather than external factors such as device recalibration.
\end{itemize}

Apart from these general assumptions, for the passive SWAP attack, it is further assumed that the attacker has knowledge of the victim’s qubit positions.

\section{Methodology} 
\label{Methodolgy}
The proposed SWAP-based side-channel attack can be executed stealthily from a distance (active attack) or used to extract sensitive information from an unsuspecting victim’s circuit (passive attack). In this section, a detailed description of both attack strategies is provided below.

\begin{figure*}[t]
    \centering
    \includegraphics[width=\linewidth]{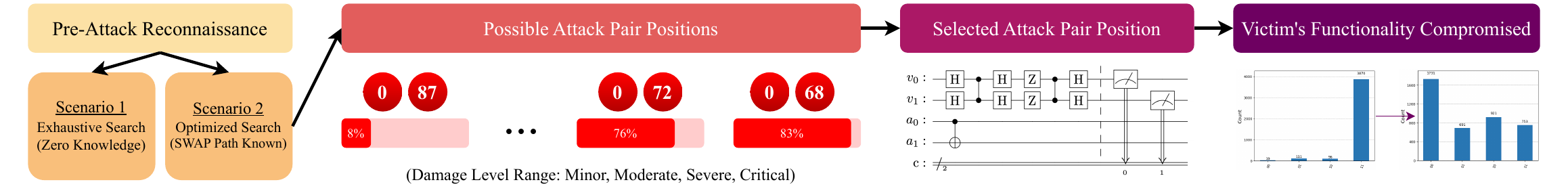}
    \caption{Active SWAP Attack Flow.}
    \label{fig:active_attack_flow}
\end{figure*}

\subsection{Active SWAP Attack}
In our proposed Active SWAP Attack, qubits positioned topologically far apart require SWAPs to communicate with one another. If victim qubits intersect with the SWAP path taken by the two attacker qubits, the resulting outcomes will be affected with a high probability. The Active SWAP Attack consists of two key steps, as follows.

In the first step, a pre-attack reconnaissance phase is conducted to identify the available attack positions and assess their potential impact on the victim's circuit, considering the topology of the quantum device. For example, in our active attack experiment, we assume the victim is located at qubit positions $63$ and $64$. The attacker can fix one qubit at an arbitrary position (e.g., position $0$) and systematically test all other positions—excluding those occupied by the victim—while measuring the victim's output accuracy. This process helps identify potential attack pairs and assess their respective impact. Specifically, it can be categorized into two scenarios depending on the attacker's knowledge of the quantum system, as shown in Fig.~\ref{fig:active_attack_flow}.
\begin{itemize}
\item \textbf{Scenario 1.} The attacker has no prior knowledge of the quantum system. Consequently, a comprehensive pre-attack reconnaissance is necessary to identify all possible attack pairs, making this approach computationally expensive. 
\item \textbf{Scenario 2.} The attacker is aware of the SWAP paths taken by qubits in the specific quantum machine. This information significantly reduces the search space, as the attacker no longer needs to evaluate all possible qubit pairs but only those intersecting the specific SWAP paths.
\end{itemize}

In the second step, the attacker can inflict damage on the victim circuit by choosing potential attack positions based on the information from the first step. The active SWAP attack can induce various levels of intensities categorized depending upon the victim's output accuracy: Minor ($20-40\%$), Moderate ($40-60\%$), Severe ($60-80\%$), Critical ($>80\%$). 
However, if the deviation in attack accuracy is less than $20\%$, it becomes very difficult to determine whether the effect is caused by the SWAP attack or inherent machine noise. Therefore, we classify it as "No Attack."
The details of the second step is detailed as follows:

\begin{enumerate}
\item Based on the list of potential attack positions identified in the first step, the attacker can estimate the expected impact of each attack. Specifically, minor attacks inflict little damage while remaining inconspicuous, whereas severe and critical attacks cause damage, leading to erroneous results upon execution.
\item Once a specific attack position is chosen, the attacker executes their circuit concurrently with the victim’s circuit at the designated attack positions.
\item Upon execution, the SWAP connectivity between the two attacker qubits induces crosstalk noise, thereby compromising the functionality of the victim circuit.
\end{enumerate}

To further illustrate the importance of the SWAP path, another active attack can be demonstrated, wherein the attacker remains in a fixed position, and the attack is triggered when the victim circuit intersects the SWAP path connecting the attacker's qubits. The details of this attack are discussed in Section~\ref{Results}.

Alternatively, the utility of SWAP attack extends beyond active attack, including passive side-channel attacks exploiting the similar vulnerability.

\begin{figure*}[t]
    \centering
    \includegraphics[width=\linewidth]{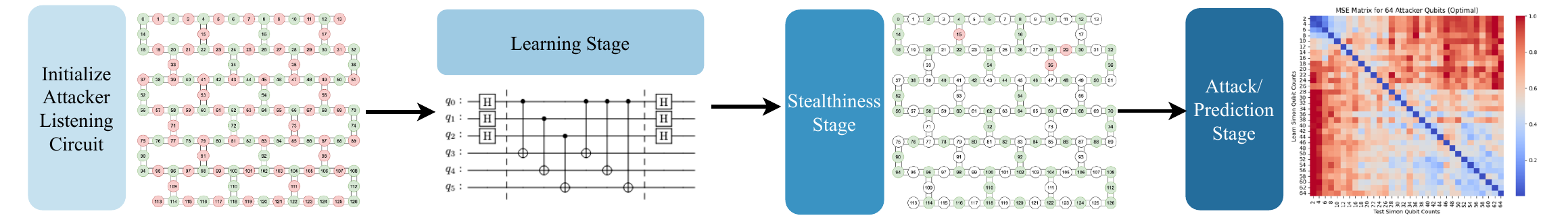}
    \caption{Passive SWAP Attack Flow.}
    \label{fig:passive_attack_flow}
\end{figure*}
 
\subsection{Passive SWAP Attack}
In a passive SWAP attack, an attacker observes the crosstalk noise generated during the execution of the victim's circuit, which can be exploited to infer sensitive information about the circuit, including its exact output and size.

The workflow of this attack is outlined in Fig.~\ref{fig:passive_attack_flow}.
\begin{itemize}
\item \textbf{Initialization.} 
First, a listening circuit with a well-defined expected output is required. In this study, we employ an empty listening circuit that consistently produces an output of all zeros. This setup is crucial as it ensures that any observed crosstalk noise originates solely from the victim qubit. Subsequently, we quantify the crosstalk effect by summing up the number of 1s measured for each listening qubit, thereby generating a crosstalk signature.
\item \textbf{Learning stage.}
At this stage, the attacker uses the listening circuit to generate different possible configurations of the victim circuit to obtain their respective unique noise signatures. This process can be divided into two main tasks: (a) victim qubit allocation and (b) noise signature collection.

During qubit allocation, the victim’s qubits are evenly distributed across the quantum device to maximize SWAP path overlap with the attacker’s listening qubits. For instance, when predicting a hidden shift length ranging from $1$ to $32$ bits, the victim’s circuit requires up to 64 qubits, allocated at even positions ($0, 2, 4, \ldots, 126$), while the attacker’s qubits occupy the odd positions ($1, 3, 5, \ldots, 125$). Consider another instance, where the attacker predicts the exact output of a fixed length, say $7$ bit, the victim’s 14 qubits are placed at strategically spaced positions (e.g., $0, 9, 19, \ldots, 126$) to maintain an even spread. The remaining qubits are assigned to the attacker’s circuit.

Finally, for noise signature collection, the unique noise signatures corresponding to the previously described victim circuit configurations are systematically evaluated. For example, when predicting the length of the hidden shift, there are $32$ distinct signatures, each corresponding to a length ranging from $1$ to $32$ bits. Similarly, when predicting the exact bit patterns of a hidden shift of size $7$ bits, there are $2^7=128$ possible values, resulting in $128$ unique noise signatures. All collected noise signatures are recorded and stored as the Signature Dataset for further analysis.

\item \textbf{Stealthiness stage.} 
By default, revealing sensitive information with maximum accuracy and confidence requires using all available qubits to construct the listening circuit. However, if the attacker prioritizes stealth to avoid detection, reducing the circuit size becomes essential. In this case, the attacker may remove qubits with minimal impact on crosstalk signatures. 

In the attacker's listening circuit, each qubit exhibits varying degrees of deviation in noise from the obtained signatures. Qubits with significant deviations are considered effective in the prediction stage of the attack, whereas those with minimal or no deviation are deemed ineffective. Based on this, the following three qubit selection strategies are proposed for the attacker's listening circuit: Optimal, Default, and Non-Optimal.
\begin{itemize}
    \item \textit{Optimal Strategy}: Qubits are ranked based on their quality and selected in descending order, from the most to the least effective, for the prediction stage.
    \item \textit{Default Strategy}: Qubits are selected according to their topological index on the quantum device, without consideration of their effectiveness in the prediction stage.
    \item \textit{Non-optimal Strategy}: Qubits are ranked in ascending order of quality, from the least to the most effective, for the prediction stage.
\end{itemize}

\item \textbf{Attack stage.} 
During the actual attack, the victim and attacker share the same quantum device. The victim executes their circuit while the attacker simultaneously runs a listening circuit on specific qubits selected during the Stealthiness Stage to capture the crosstalk signature of the victim circuit.
\item \textbf{Prediction stage.} 
In the final stage, we compare the victim's crosstalk signature with the Signature Dataset, expecting it to closely match the training signature for the victim's circuit size. The closest match is identified using the mean square error (MSE) formula, as provided in Equation~\ref{formula:mse}.
\begin{equation}
MSE = \frac{1}{n} \sum_{i=1}^{n} (y_i - \hat{y}_i)^2
\label{formula:mse}
\end{equation}
The MSE between each entry in the Signature Dataset and the victim's crosstalk signature is ranked, with the lowest MSE representing the best prediction. 
\end{itemize}

In summary, the proposed passive SWAP attack effectively extracts sensitive information from other quantum circuits sharing common quantum resources. This attack primarily exploits improper qubit allocation on quantum cloud platforms, resulting in distinct crosstalk patterns observed during measurements on the attacker's listening qubits. Therefore, designing a more secure qubit allocation protocol is essential as an effective defense strategy against this attack.

\section{Results and Discussions} ~\label{Results}
In this section, we provide detailed descriptions of our experiments as follows.

\subsection{Experimental Setup}
For all the experiments in this paper, we use \textit{ibm\_brisbane}, a $127$-qubit cloud-based quantum processor provided by IBM.\footnote{Earlier experiments were conducted on \textit{ibm\_kyoto}, which has since been deprecated.} To ensure consistency in SWAP paths between distant qubits, we fix the \textit{pass manager}'s randomized transpilation heuristic to \textit{seed\_transpile} $= 0$. Additionally, to minimize variations due to quantum hardware calibration, all experiments were queued back-to-back, maximizing the chances of execution within the same calibration window.

In active SWAP attack experiments, we evaluate the attack's impact using the output accuracy metric, $Acc_0$. This metric is defined as the ratio of shots producing the expected result to the total number of observed shots, which is set to $4096$ by default. In the absence of any adversarial attack, $Acc_0$ remains slightly below $1$ due to inherent machine noise.

In passive SWAP attack experiments, we use two metrics to evaluate the performance of our attack: accuracy ($Acc_1$) and Confidence.
For accuracy, as defined in Formula \ref{formula:prediction_accuracy}, 
$n$ denotes the number of unique signatures in our Signature Dataset, while $i$ represents the zero-indexed rank of the actual victim circuit configuration among all signatures sorted by the lowest mean squared error (MSE). For confidence, we calculate the difference in normalized MSE values between lowest $MSE_{l}$ and 2nd lowest signature $MSE_{s}$ as shown in Formula~\ref{formula:prediction_confi}. 
\begin{equation}
Acc_1 = \frac{n - i}{n} 
\label{formula:prediction_accuracy}
\end{equation}
\begin{equation}
Confidence = MSE_{l} - MSE_{s}
\label{formula:prediction_confi}
\end{equation}

\subsection{Experiment 1: Active SWAP Attack}

\begin{figure}[ht!]
    \centering
    \includegraphics[width=0.99\linewidth]{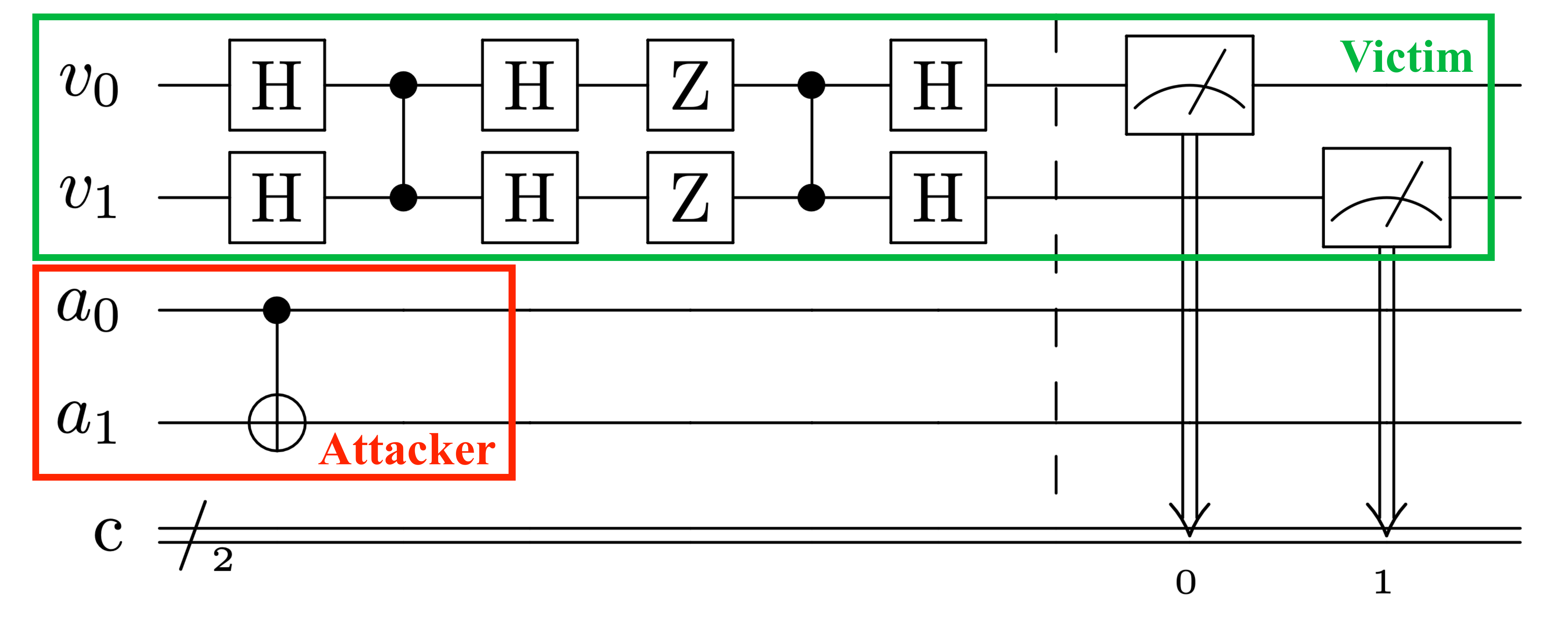}
    \caption{Attack Configuration in Experiment 1. Here, victim running Grover's circuit on qubits $v_0, v_1$, with $\ket{11}$ being the desired output. Attacker is running the CNOT gate on the qubits $a_0, a_1$.}
    \label{fig:grover}
\end{figure}

As illustrated in Fig.~\ref{fig:grover}, we assume the victim is executing Grover’s Algorithm~\cite{grover96} in Experiment 1, which is a well-known quantum algorithm in cryptography. Meanwhile, the lightweight attacker executes a single CNOT gate. Importantly, this proposed attack design improves upon previous works~\cite{ash2020analysis, deshpande2023design, harper2024crosstalk}, which rely on multiple successive CNOT gates to carry out the attack.

In Section~\ref{Methodolgy}, two attack scenarios of active SWAP attack are already defined in details. Scenario 1 requires comprehensive pre-attack reconnaissance, where the attacker systematically explores all potential attack positions. In contrast, Scenario 2 assumes the attacker has prior knowledge of the physical quantum hardware and the SWAP paths between qubits, hence significantly reduces the search space.\footnote{The attacker only needs to iterate through pairs of qubits connected by SWAP paths that intersect the victim's circuit.}

\begin{figure}[ht!]
    \centering
    \includegraphics[width=1.0\linewidth]{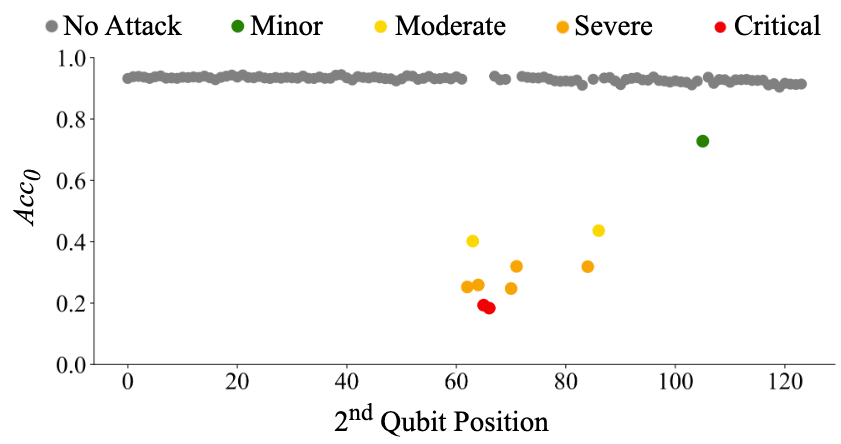}
    \caption{The Output Accuracy, $Acc_0$ of Victim Circuit Estimation Across Different Qubit Positions.}
    \label{fig:active_attack_output_accuracy}
\end{figure}

Here, we exhaustively evaluate all possible attack pairs under Scenario 1. The results of the pre-attack reconnaissance are illustrated in Fig.~\ref{fig:active_attack_output_accuracy}. From this figure, it is evident that different attack positions can induce varying levels of attack severity. Specifically, if $Acc_0$ drops below $0.5$—indicating that the correct output is observed in less than $50\%$ of cases—the attack is classified as moderate ($40-60\%$), severe ($60-80\%$), or critical ($>80\%$) depending on the amount of deviation in the victim's expected output. Conversely, if the $Acc_0$ remains above $0.5$, the attack is classified as moderate ($40-60\%$), minor ($20-40\%$), or no attack ($0-20\%$). According to these results, an attack strategy list is derived as Table \ref{tab1}.

\vspace{10pt}

\begin{table}[h]
\begin{center}
\begin{tabular}{| c | c | c | c|}
\hline
\textbf{2nd Qubit Position}  & \textbf{\textbf{Acc\_0}} &\textbf{ Deviation ($\%$)} & \textbf{Severity}\\
\hline
\rowcolor{Severerow}
{65} & {0.25269} & ${74.73\%}$ & {Severe}\\
\hline
\rowcolor{Moderaterow}
{66} & {0.40210} & ${59.79\%}$ & {Moderate}\\
\hline
\rowcolor{Severerow}
{67} & {0.25928} & ${74.07\%}$ & {Severe}\\
\hline
\rowcolor{Criticalrow}
{68} & {0.19336} & ${80.66\%}$ & {Critical}\\
\hline
\rowcolor{Criticalrow}
{69} & {0.18384} & ${81.62\%}$ & {Critical}\\
\hline
\rowcolor{Severerow}
{73} & {0.24756} & ${75.24\%}$ & {Severe}\\
\hline
\rowcolor{Severerow}
{74} & {0.31982} & ${68.02\%}$ & {Severe}\\
\hline
\rowcolor{Severerow}
{87} & {0.31885} & ${68.12\%}$ & {Severe}\\
\hline
\rowcolor{Moderaterow}
{89} & {0.43628} & ${56.37\%}$ & {Moderate}\\
\hline
\rowcolor{Minorrow}
{108} & {0.72803} & ${27.20\%}$ & {Minor}\\\hline
\end{tabular}
\caption{Severity of Attack Intensity Depending on Qubit Positions. Here, we assumed that the first qubit of the attacker is fixed at the $0$-th position, and the attacker varies the second qubit to estimate the output accuracy.}
\label{tab1}
\end{center}
\end{table}

This strategy list provides the attacker with options for varying degrees of disruption. The attacker may choose a high-intensity attack, which significantly corrupts the victim’s output but risks detection, or a low-intensity attack, which induces subtle errors that are harder to detect on a quantum machine, enabling stealthier attacks.




\subsection{Experiment 2: Impact of SWAP Path on Active Attack}
\begin{figure}[h]
    \centering
    \includegraphics[width=1.0\linewidth]{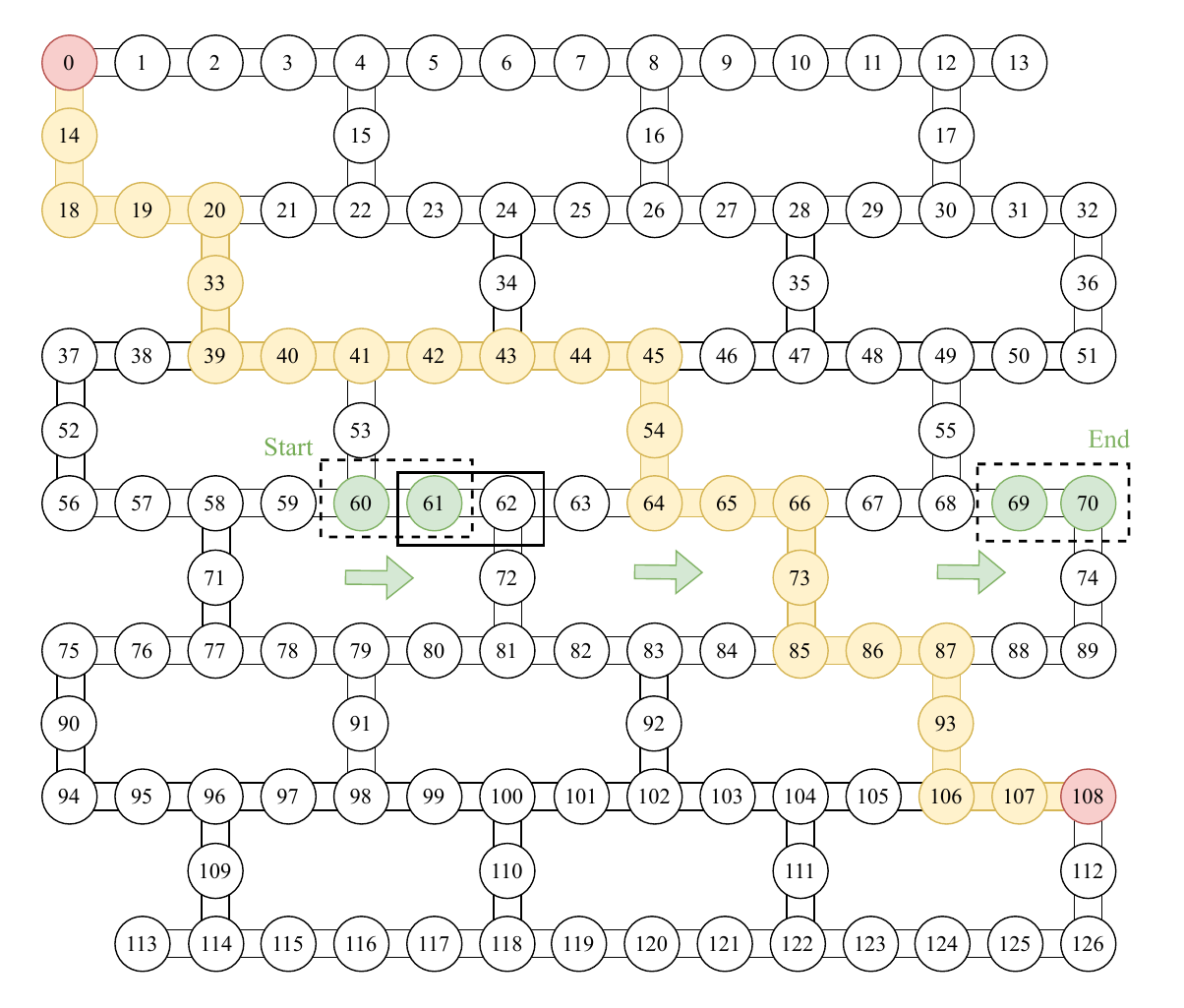}
    \caption{Attack Configuration in Experiment 2. Specifically, it illustrates the SWAP path between the attacker's qubits at positions 0 and 108.}
    \label{fig:active_attack_SWAP_path}
\end{figure}
In this experiment, the attacker is positioned at qubits 0 and 108, while the victim's circuit is placed at various positions, starting from position pair $(60,61)$ and shifting one qubit to the right until reaching $(69,70)$. As illustrated in Fig.~\ref{fig:active_attack_SWAP_path}, four position pairs intersect the attacker's SWAP path, while the other six do not.

\begin{figure}[ht]
    \centering
    \includegraphics[width=0.86\linewidth]{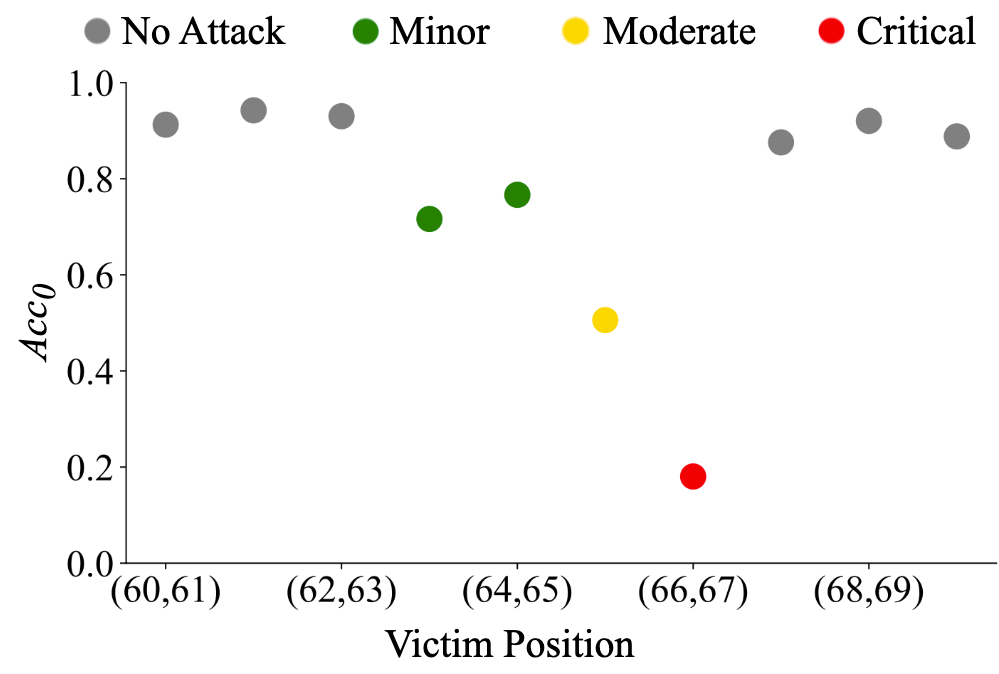}
    \caption{The Output Accuracy, $Acc_0$ for Different Positions of Victim Circuit with a Fixed Attacker Position.}
    \label{fig:active_attack_fix_attacker_vary_victim_graph}
\end{figure}

The attack results, presented in Fig.~\ref{fig:active_attack_fix_attacker_vary_victim_graph} and Table~\ref{tab2}, show no significant deviation from the expected output when the victim's qubits do not intersect with the attacker's SWAP path, as observed for qubit pairs $(60, 61)$, $(61, 62)$, $(62, 63)$, $(67, 68)$, $(68, 69)$, and $(69, 70)$. However, when the victim's qubits do intersect the SWAP path, an obvious increase in deviation is observed. The highest deviation, $81.96\%$, occurs at qubit pair position $(66, 67)$, where the output accuracy drops to $0.18042$, leading to an critical erroneous result.

In brief, this experiment clearly highlight the critical role of the SWAP path in the effectiveness of the proposed active attack.


\begin{table*}[ht]
\begin{center}
\begin{tabular}{| c | c | c | c | c | c |}
\hline
\textbf{Victim's 1st Qubit} & \textbf{Victim's 2nd Qubit} & \textbf{Attacker's SWAP Path Intersection} & \textbf{Acc\_0} &\textbf{ Deviation ($\%$)} & \textbf{Severity}\\
\hline
\rowcolor{noattackrow}
60 & 61 & FALSE & $0.91260$ & $8.74\%$ & No Attack\\
\hline
\rowcolor{noattackrow}
61 & 62 & FALSE & 0.94238 & $5.76\%$ & No Attack\\
\hline
\rowcolor{noattackrow}
62 & 63 & FALSE & $0.93018$ & $6.98\%$ & No Attack\\
\hline
\rowcolor{Minorrow}
63 & 64 & TRUE & $0.71631$ & $28.37\%$ & Minor\\
\hline
\rowcolor{Minorrow}
64 & 65 & TRUE & $0.76660$ & $23.34\%$ & Minor\\
\hline
\rowcolor{Moderaterow}
65 & 66 & TRUE & $0.50586$ & $49.41\%$ & Moderate\\
\hline
\rowcolor{Criticalrow}
66 & 67 & TRUE & $0.18042$ & $81.96\%$ & Critical\\
\hline
\rowcolor{noattackrow}
67 & 68 & FALSE & $0.87573$ & $12.43\%$ & No Attack\\
\hline
\rowcolor{noattackrow}
68 & 69 & FALSE & $0.92041$ & $7.96\%$ & No Attack\\
\hline
\rowcolor{noattackrow}
69 & 70 & FALSE & $0.88818$ & $11.18\%$ & No Attack\\\hline
\end{tabular}
\caption{Attack Severity with Fixed Attacker and Varying Victim Qubit Positions. In this experiment, $Acc_0$ decreases when the victim circuit intersects the attacker's SWAP path.}
\label{tab2}
\end{center}
\end{table*}


\subsection{Experiment 3: Passive SWAP Attack for Circuit Size}

\begin{figure}[h]
    \centering
    \includegraphics[width=0.9\linewidth]{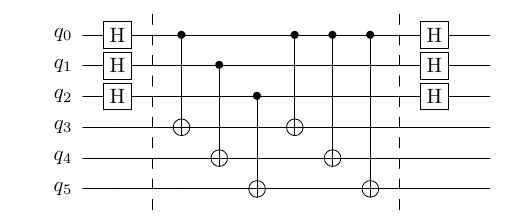}
    \caption{Simon's Algorithm with a Hidden Shift $s = 111$.}
    \label{fig:simon_111} 
\end{figure}

For the passive SWAP attacks in Experiments 3 and 4, the scalability of the victim’s circuit is crucial. Therefore, Simon's algorithm~\cite{simon} is used as the victim circuit (Fig.~\ref{fig:simon_111}) to avoid the exponential scaling of qubits observed in Grover's algorithm.

In this experiment, we focus on predicting the size of the victim’s circuit. A series of victim circuits were tested, with sizes ranging from $2$ to $64$ qubits in increments of $2$ qubits. Successfully predicting the victim’s circuit size enables the attacker to infer the length of the hidden shift $s$, which is half the size of the victim’s circuit.

To maximize the distance between qubits, and given that the maximum victim circuit size is $64$ qubits, the victim’s circuit may occupy qubit positions $0,2,4, \ldots , 124, 126$. This placement ensures that at least one qubit remains available for the attacker to monitor. For the attack circuit, the maximum size is $127-64 = 63$ qubits. 

The attack process follows the same workflow described in Section~\ref{Methodolgy}.
During initialization, the attacker sets up a $63$-qubit listening circuit, expecting an output of all zeros. Any deviation from this expected result is attributed to distant crosstalk noise between victim qubits.

\begin{figure}[ht!]
\centering
  \includegraphics[width=\linewidth]{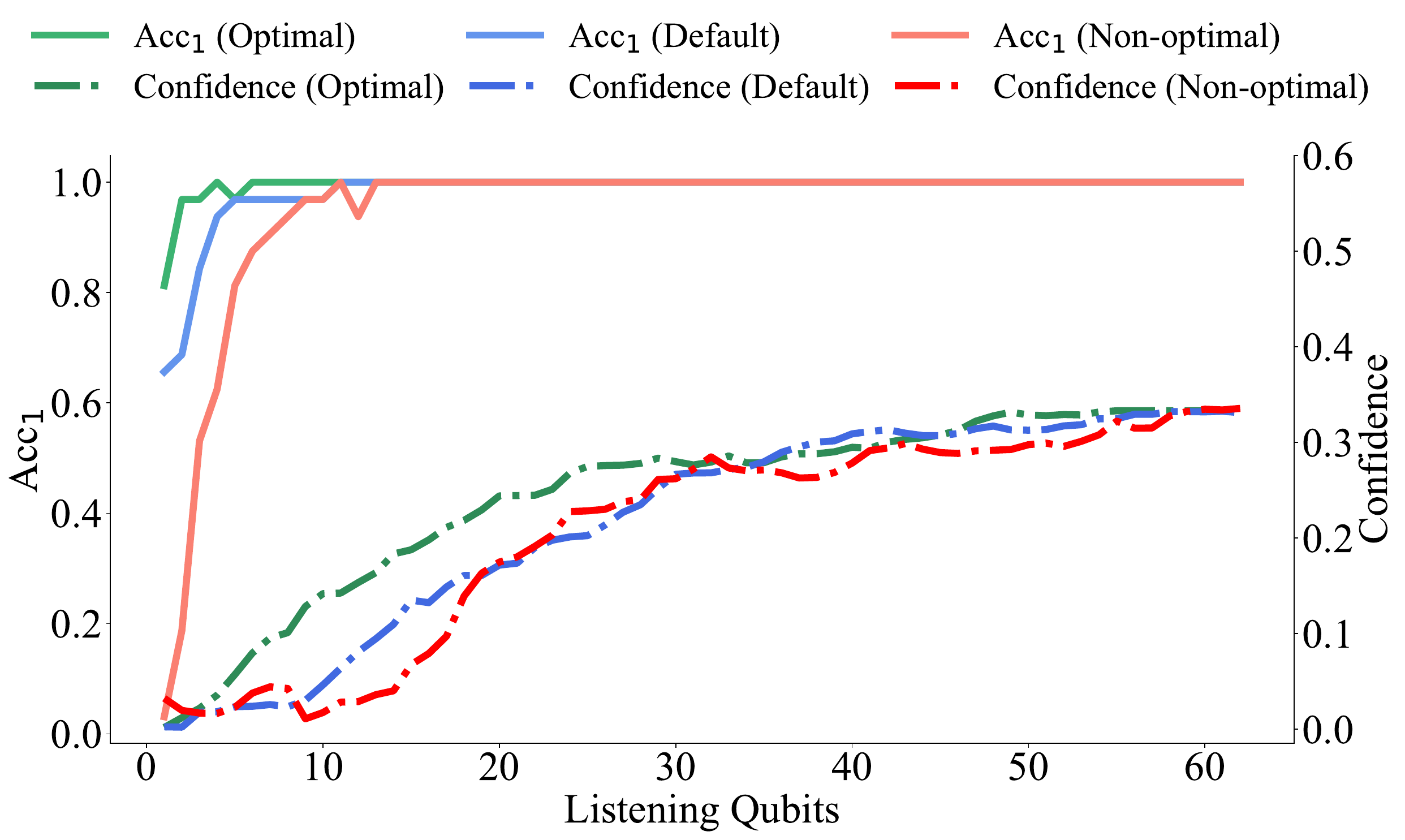}
\caption{Tradeoff Between Listening Size, $Acc_1$, and Confidence in Experiment 3.\label{fig:passive_attack_selecting_qubits}}
\end{figure}

In the learning phase, the attacker iteratively acquires and records unique crosstalk signatures corresponding to all $32$ possible circuit sizes, compiling them into a Signature Dataset. By analyzing the collected signatures, the attacker may opt to further reduce their listening circuit for a stealthier attack, to minimize detection. As illustrated in Fig.~\ref{fig:passive_attack_selecting_qubits}, selecting a minimum of 4 optimally placed qubits is sufficient to achieve $100\%$ accuracy in side-channel prediction.

Once the attacker's listening circuit size has been determined, the attacker can proceed with executing the attack on the actual quantum device along with the victim's circuit. After the victim's circuit has completed its execution, the obtained crosstalk signature is compared against all entries in the Signature Dataset. The closest match, based on the lowest MSE, is considered the best prediction.

\begin{figure*}[ht!]
\centering
\begin{subfigure}{.3\textwidth}
  \centering
  \includegraphics[width=0.95\linewidth]{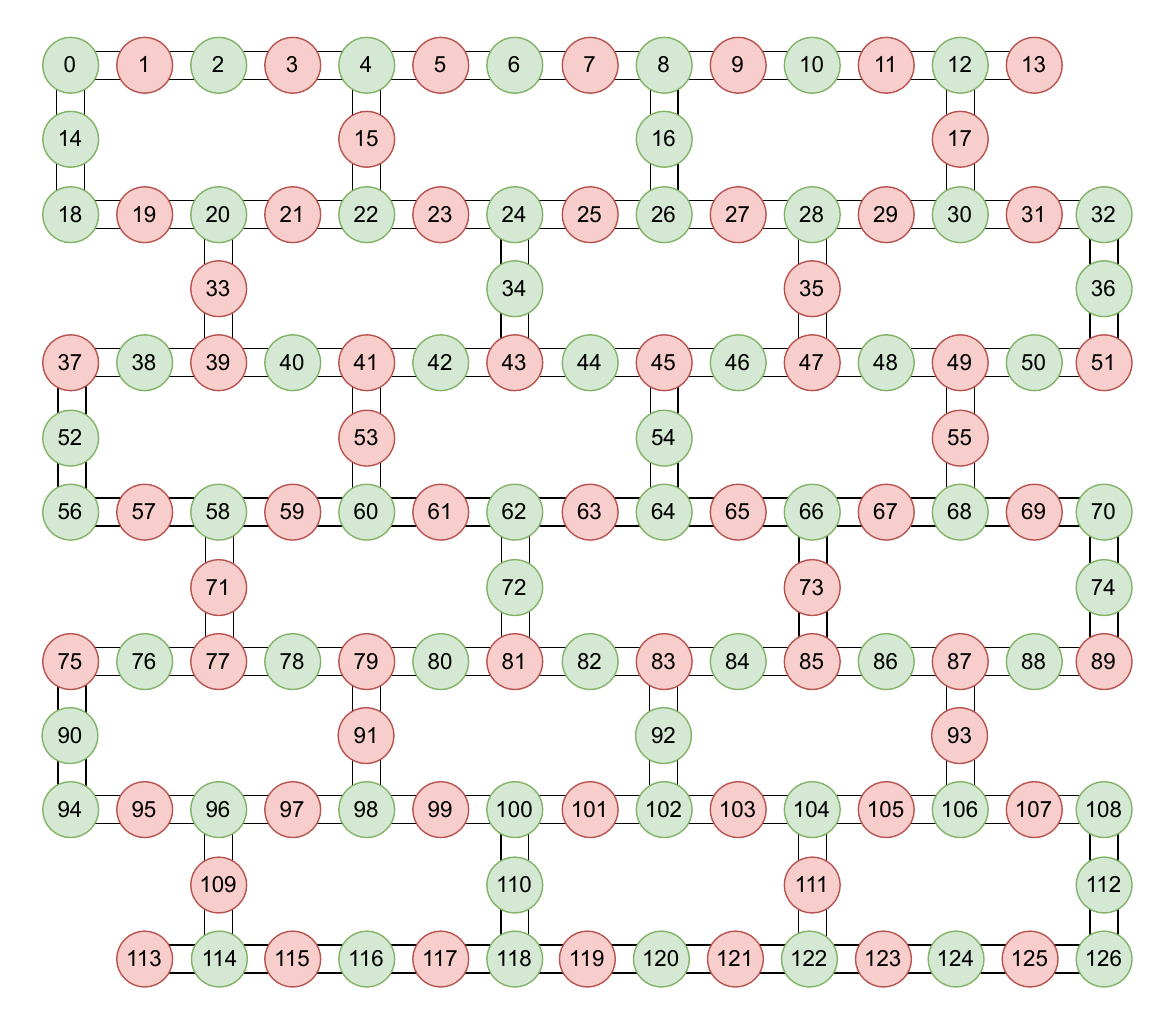}
  \caption{\#Attacker Qubits: $63$.}
\label{fig:side_channel_optimal_all_qubits_layout.pdf}
\end{subfigure}%
\begin{subfigure}{.3\textwidth}
  \centering
  \includegraphics[width=0.95\linewidth]{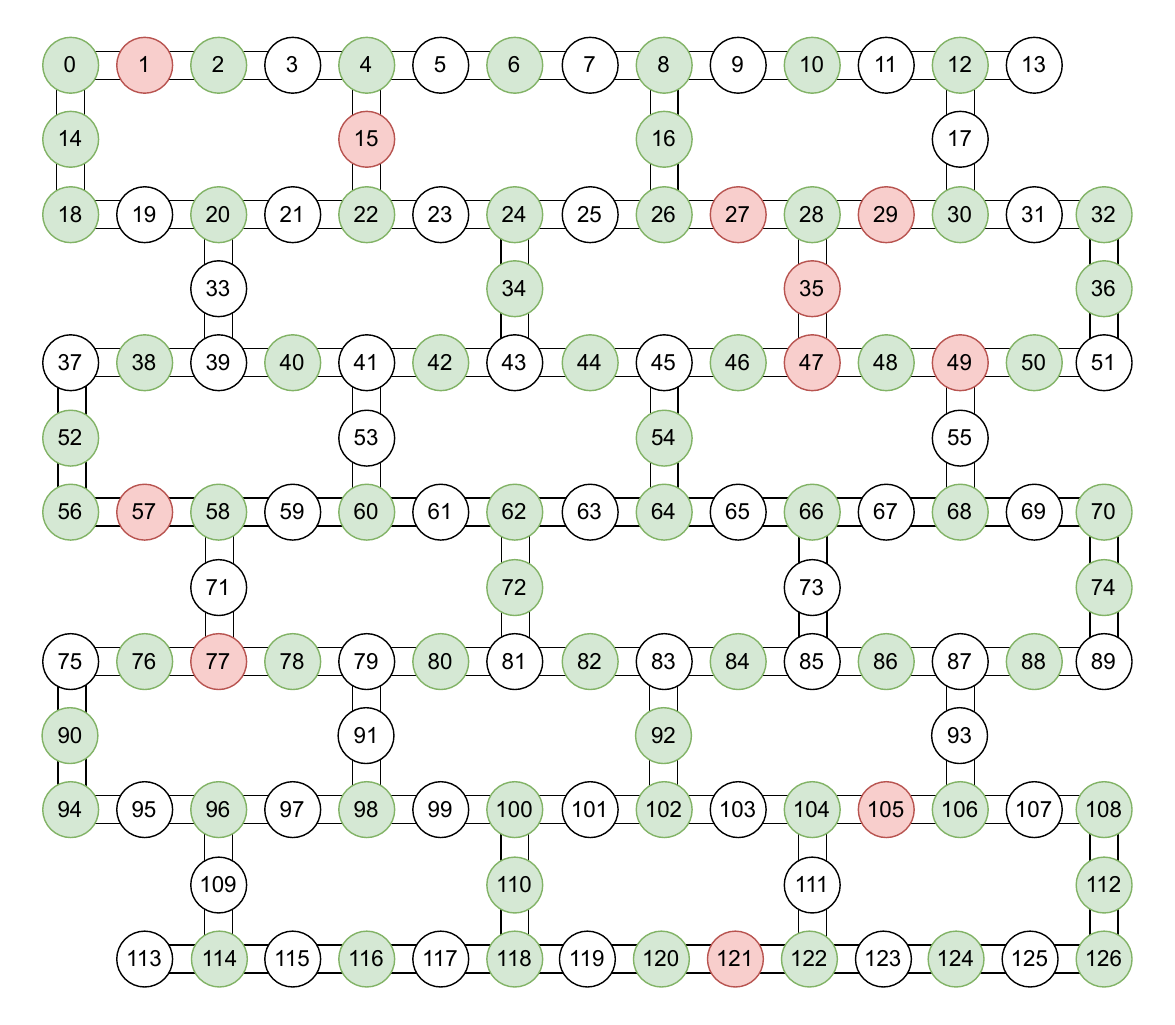}
  \caption{\#Attacker Qubits: $11$.}
\label{fig:passive_attack_predict_size_11_qubit_layout}
\end{subfigure}%
\begin{subfigure}{.3\textwidth}
  \centering
  \includegraphics[width=0.95\linewidth]{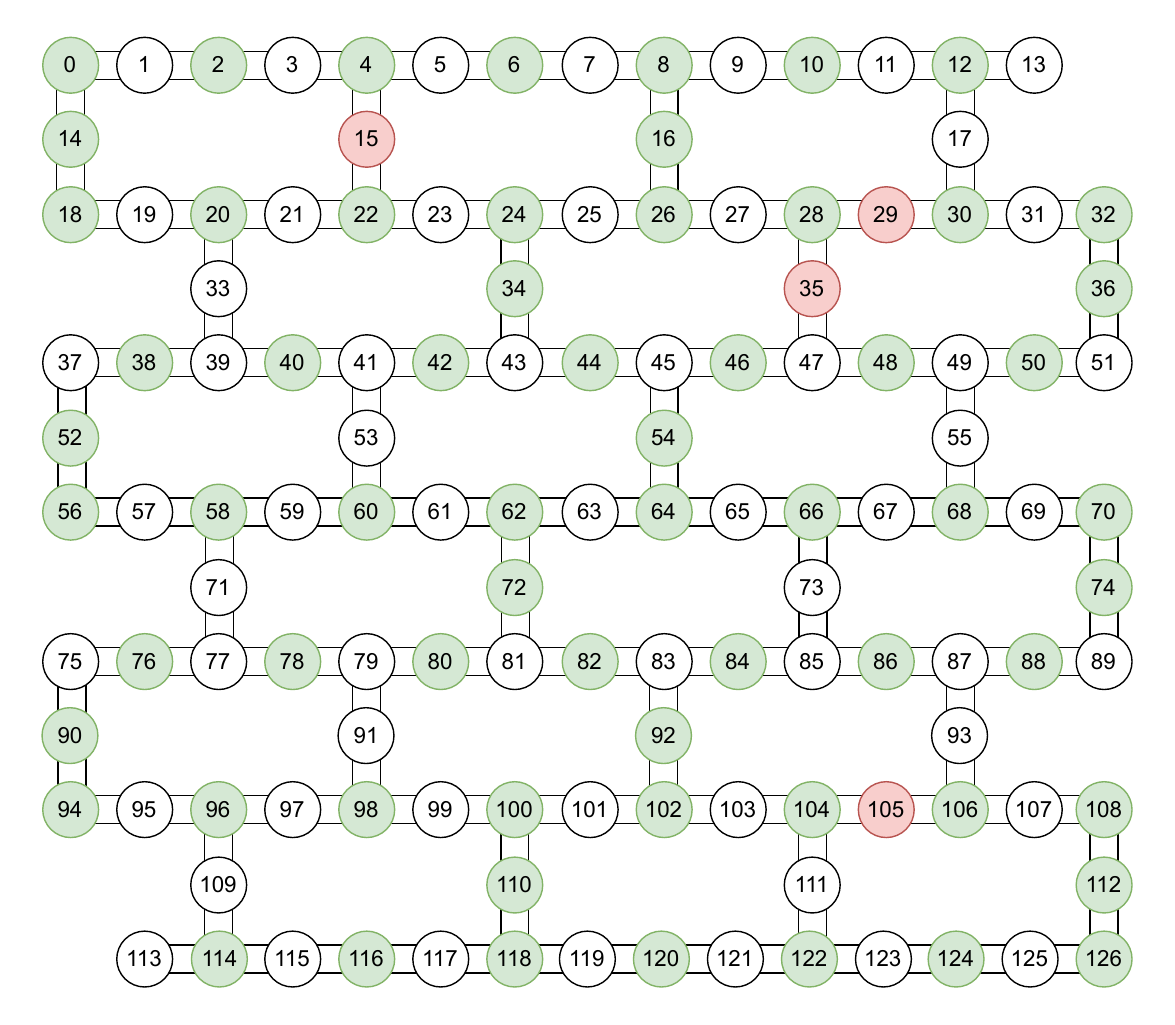}
  \caption{\#Attacker Qubits: $4$.}
\label{fig:passive_attack_predict_size_4_qubit_layout}
\end{subfigure}%
\hfill
\begin{subfigure}{.3\textwidth}
  \centering
  \includegraphics[width=.955\linewidth]{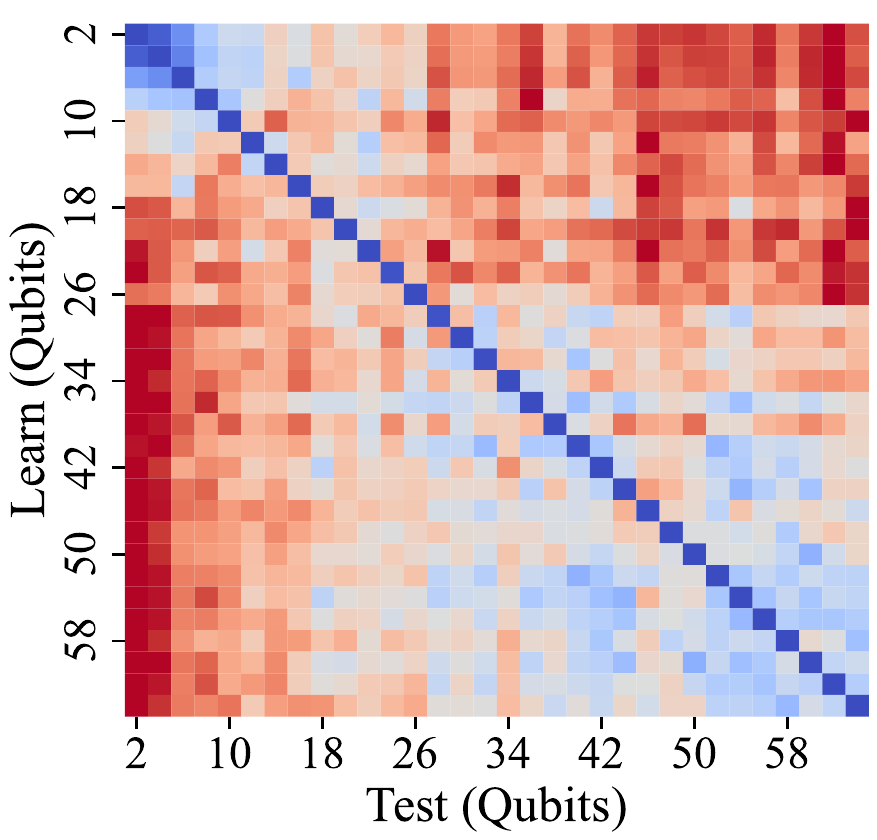}
  \caption{MSE Matrix for $63$ Qubits}
  \label{fig:passive_attack_predict_size_all_matrix}
\end{subfigure}%
\begin{subfigure}{.3\textwidth}
  \centering
  \includegraphics[width=.955\linewidth]{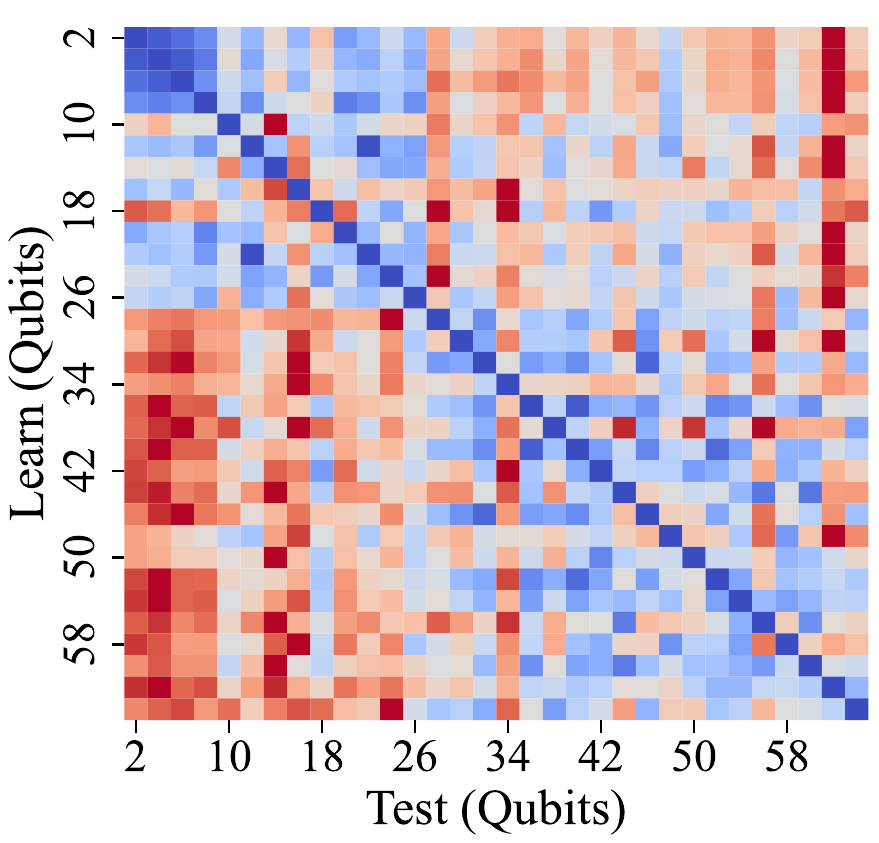}
  \caption{MSE Matrix for $11$ Qubits}
  \label{fig:passive_attack_predict_size_11_matrix}
\end{subfigure}%
\begin{subfigure}{.3\textwidth}
  \centering
  \includegraphics[width=1.13\linewidth]{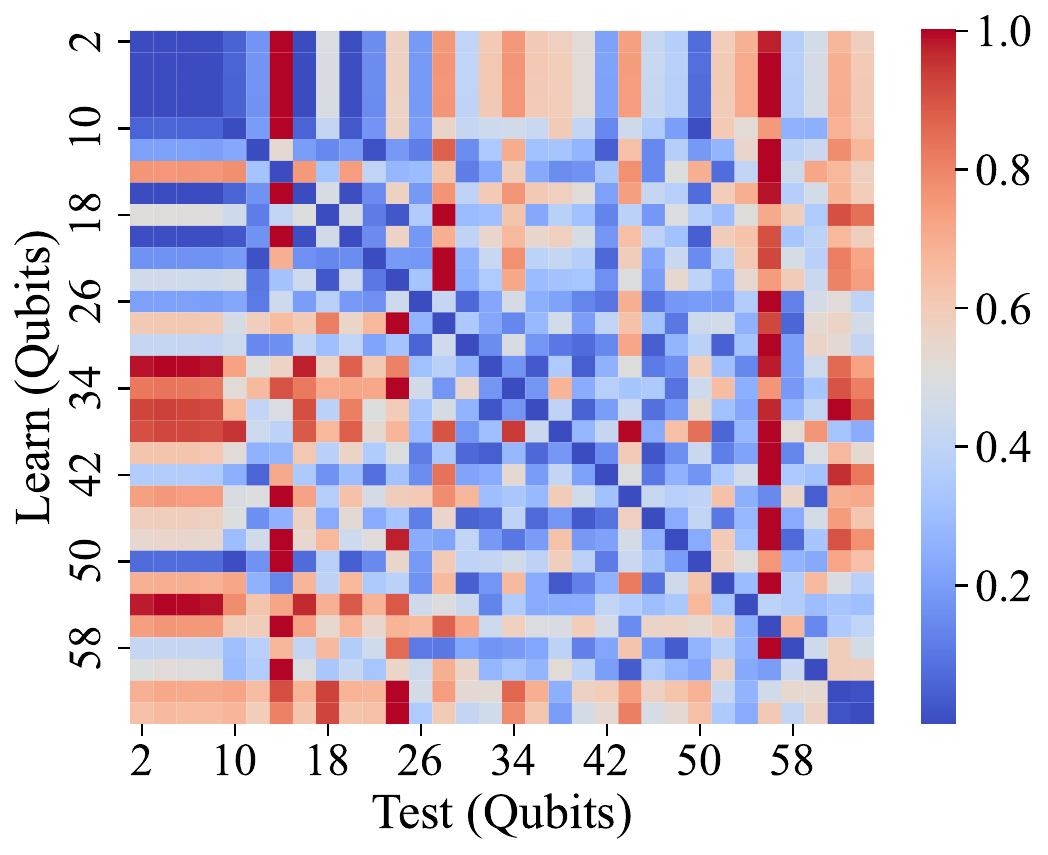}
  \caption{MSE Matrix for $4$ Qubits}
  \label{fig:passive_attack_predict_size_4_matrix}
\end{subfigure}

\caption{Attack Configuration and Results in Experiment 3.}
\label{fig:passive_predict_size_matrix}
\end{figure*}

Fig.~\ref{fig:passive_predict_size_matrix} presents the listening circuits with size $63$, $11$, and $4$ qubits, along with the corresponding MSE matrix. The results indicate that as the number of listening qubits increases, the confidence of the side-channel prediction also improves.

\begin{figure*}[hb]
\centering
\begin{subfigure}{.3\textwidth}
  \centering
  \includegraphics[width=0.95\linewidth]{passive_predict_size_optimal_11.pdf}
  \caption{Optimal}
\label{fig:passive_predict_size_optimal_11.pdf}
\end{subfigure}%
\begin{subfigure}{.3\textwidth}
  \centering
  \includegraphics[width=0.95\linewidth]{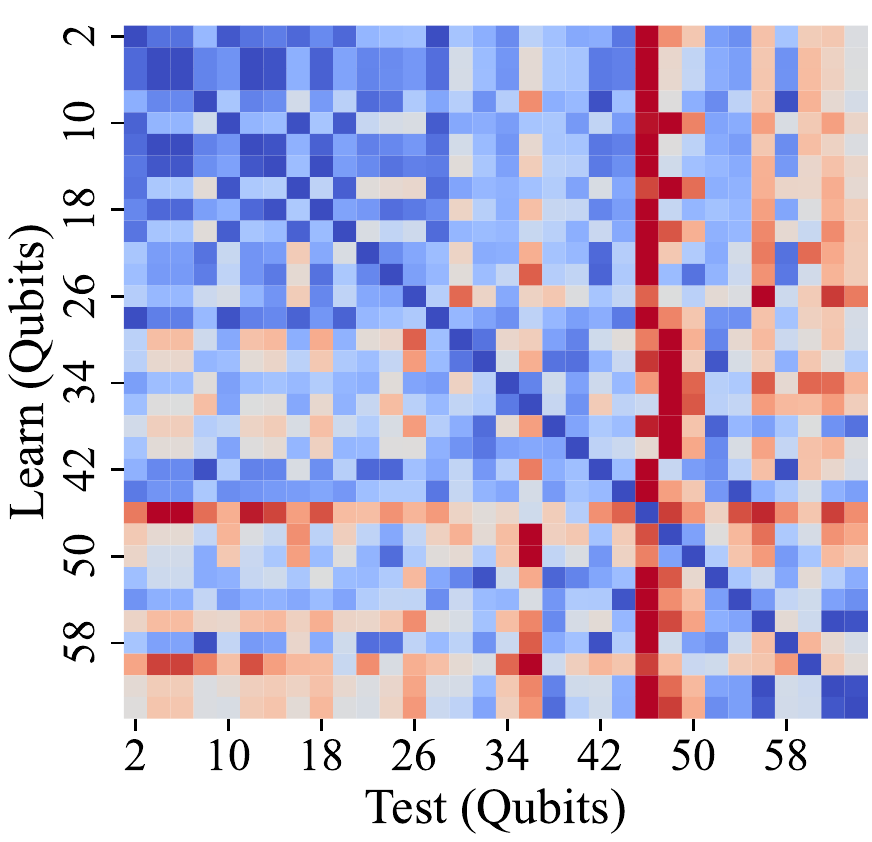}
  \caption{Default}
\label{fig:passive_predict_size_default_11}
\end{subfigure}%
\begin{subfigure}{.3\textwidth}
  \centering
  \includegraphics[width=1.12\linewidth]{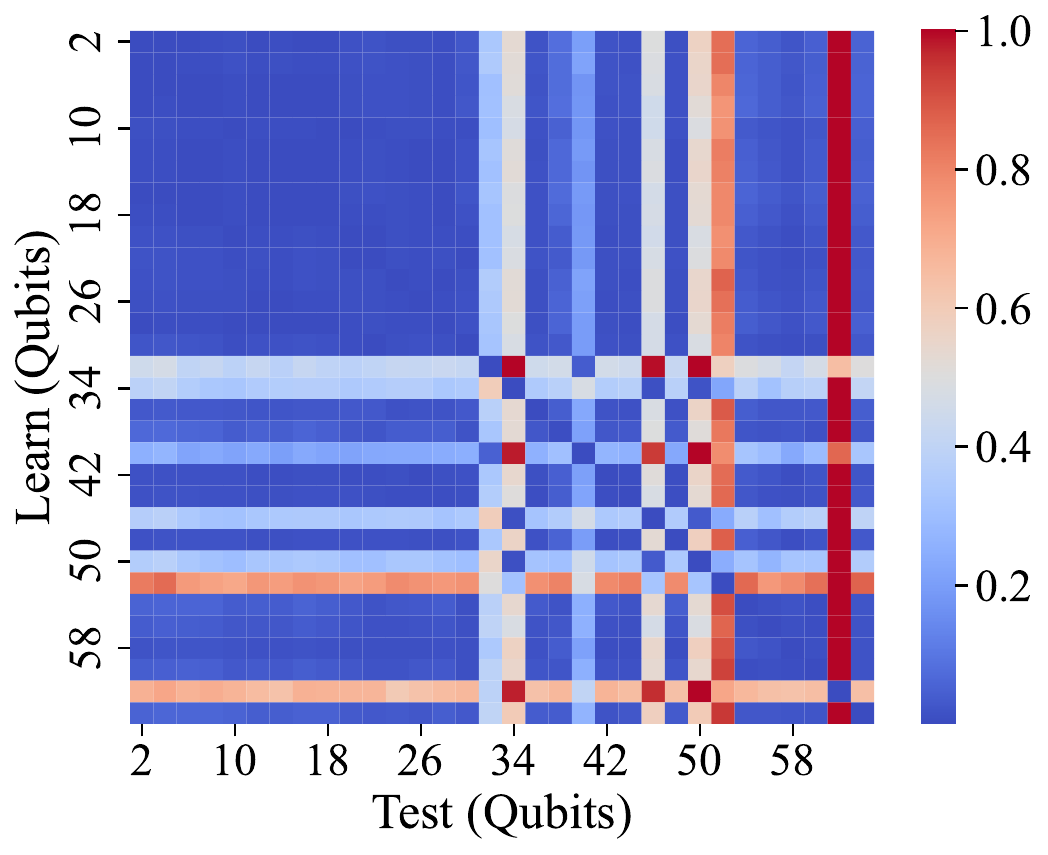}
  \caption{Non-optimal}
\label{fig:passive_predict_size_non-optimal_11.pdf}
\end{subfigure}%

\caption{MSE Matrix for Listening Circuit with $11$ Qubits.}
\label{fig:passive_attack_qubit_selection_matrix_comparison}
\end{figure*}

From the experimental results, we observe a trade-off between the number of qubits used in the attacker's circuit and the prediction accuracy of the side-channel attack. Although, as previously mentioned, the attacker requires only 4 qubits to achieve $100\%$ accuracy in side-channel prediction, the confidence in this prediction remains low. Increasing the number of qubits allows the attacker to make predictions with greater confidence, as the MSE percentage increases. This trade-off is crucial for attackers who seek high prediction accuracy while maintaining a smaller circuit, as a large listening circuit with no gates or operations other than measurement may be detected and flagged by an antivirus system on the quantum device.

\vspace{-5pt}

Beyond increasing the number of qubits in the listening circuit, the selection of qubits is also critical. Given a fixed number of 11 listening qubits, there are 3 possible qubit selection strategies: Optimal, Default, and Non-optimal. The corresponding MSE matrices are shown in Fig.~\ref{fig:passive_attack_qubit_selection_matrix_comparison}, highlight significant differences in prediction confidence despite all 3 strategies achieving $100\%$ accuracy. Among them, the non-optimal qubit selection strategy yields the lowest confidence, while the optimal qubit selection provides the highest confidence.

Thus, this trade-off grants attackers greater flexibility in their approach. They may choose to prioritize prediction accuracy and confidence or, alternatively, prioritize stealth by employing the smallest possible listening circuit that maintains high accuracy, albeit with reduced confidence.

\subsection{Experiment 4: Passive SWAP Attack for Output.}
In this experiment, the victim executes Simon’s algorithm for a $7$-bit hidden shift, requiring 14 qubits. While the size of the hidden shift is known to the attacker, its exact value remains unknown. The attacker employs a passive SWAP attack to infer the hidden shift, which consists of $2^7$ possible values.

Specifically, the victim’s qubits are allocated at positions $0, 9, 19, 29, 38, 48, 58, 67, 77, 87, 96, 106, 116$, and $126$ to ensure even distribution across the quantum device.

\begin{figure}[ht!]
\centering
  \includegraphics[width=0.99\linewidth]{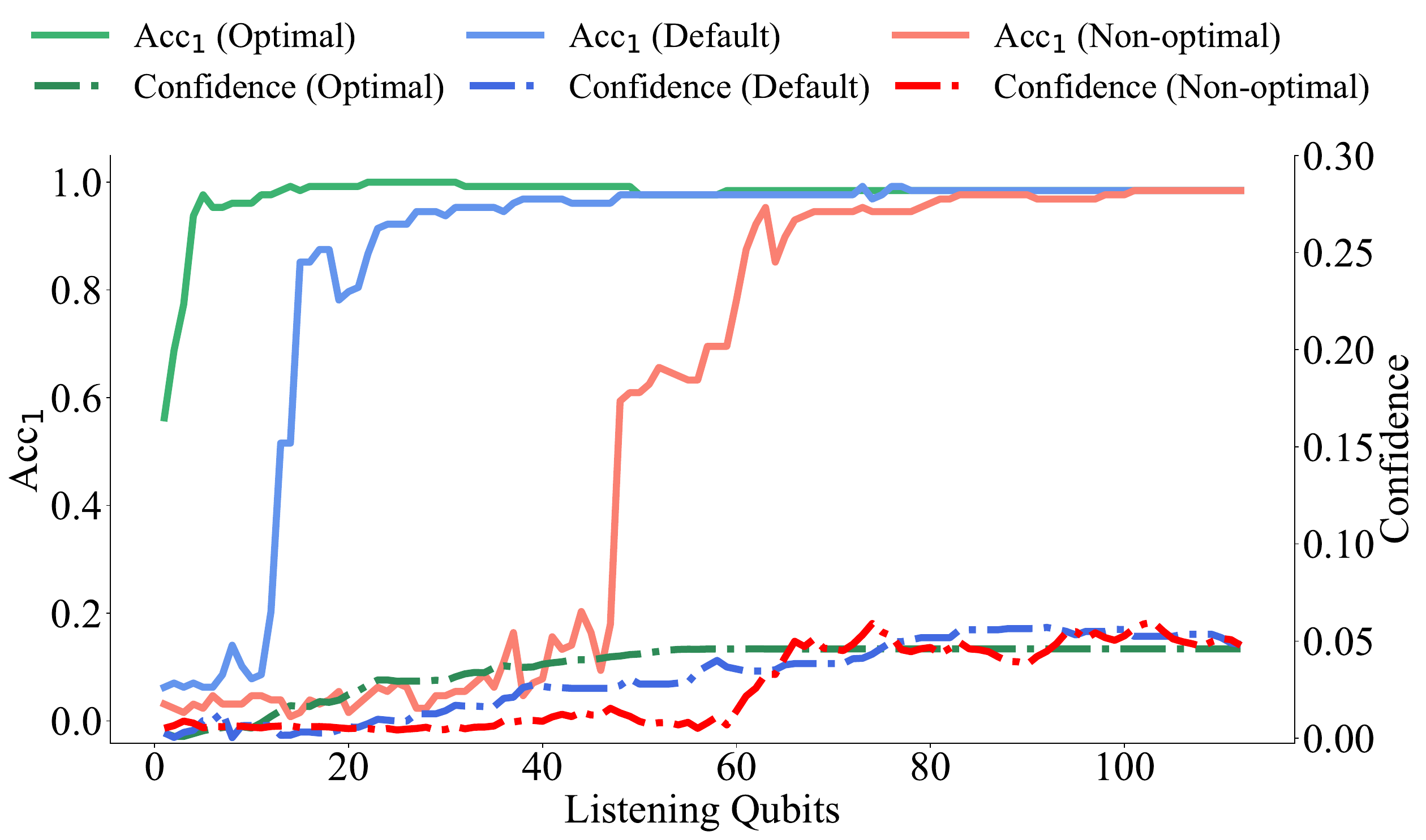}
\caption{Tradeoff Between Listening Size, $Acc_1$, and Confidence in Experiment 4.\label{fig:111}}
\end{figure}

For the attack circuit, different strategies are applied as illustrated in Fig.~\ref{fig:111}. Under the Optimal Strategy, a minimum of $22$ qubits is required to achieve $100\%$ accuracy in predicting the hidden shift bit pattern. Using the Default Strategy, at least $30$ qubits are needed to exceed $95\%$ accuracy. In contrast, the Non-optimal Strategy requires a minimum of $62$ qubits to achieve more than $95\%$ accuracy.

\begin{figure}[ht!]
    \centering
  \includegraphics[width=0.99\linewidth]{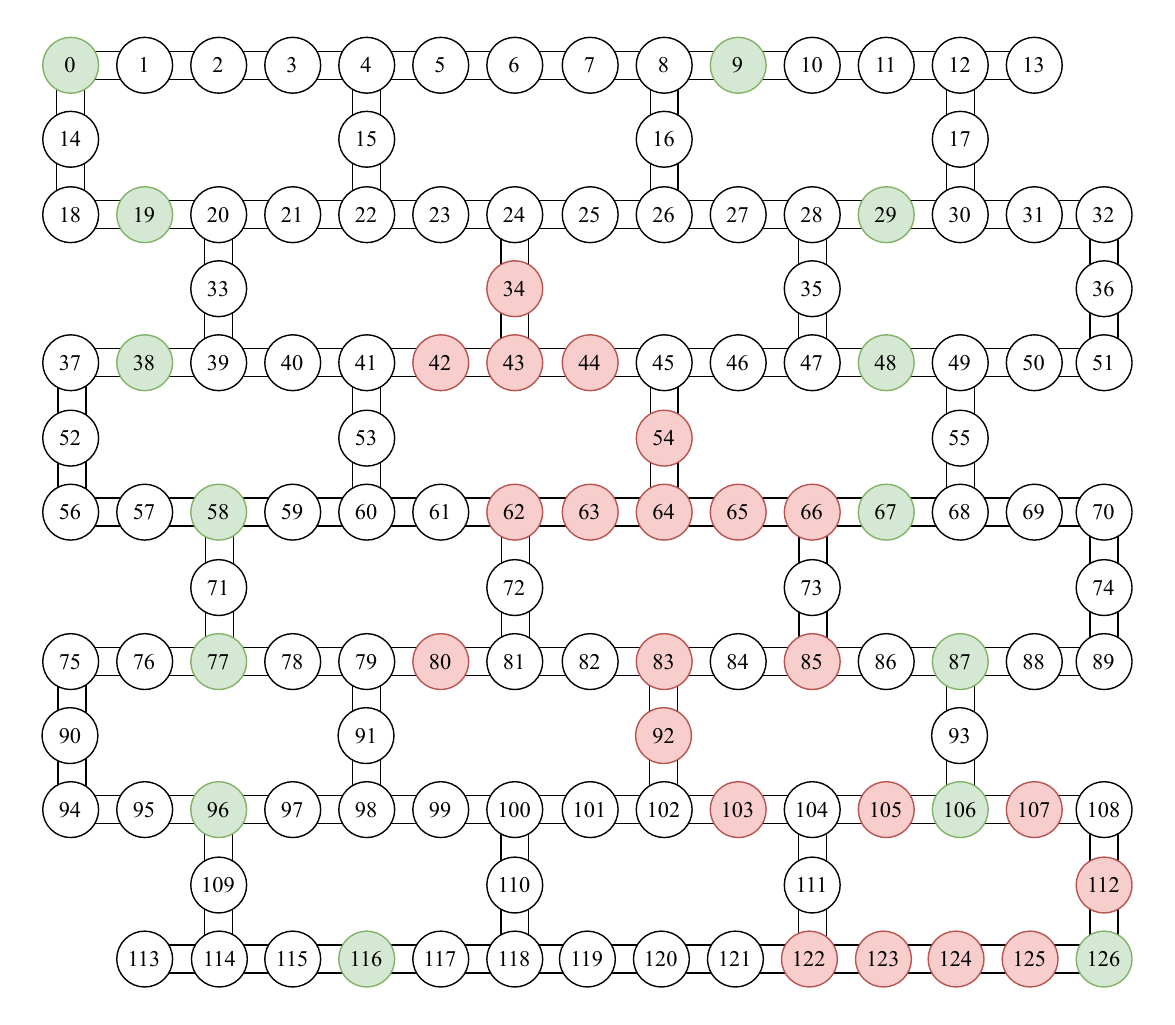}
\label{fig:passive_attack_predict_size_3_qubit_layout}
\caption{Attack Configuration in Experiment 4. Here, red qubits represent the attacker circuit (Optimal Strategy), green qubits represent the victim circuit (Even Distribution).}
\label{fig:222}
\end{figure}

After selecting the attack circuit, the attacker proceeds with the passive SWAP attack to predict the output value. With the Optimal Strategy, only $22$ qubits are necessary to achieve $100\%$ prediction accuracy, as illustrated in Fig.~\ref{fig:222}.

\begin{figure*}[ht!]
\centering
\begin{subfigure}{.3\textwidth}
  \centering
  \includegraphics[width=0.953\linewidth]{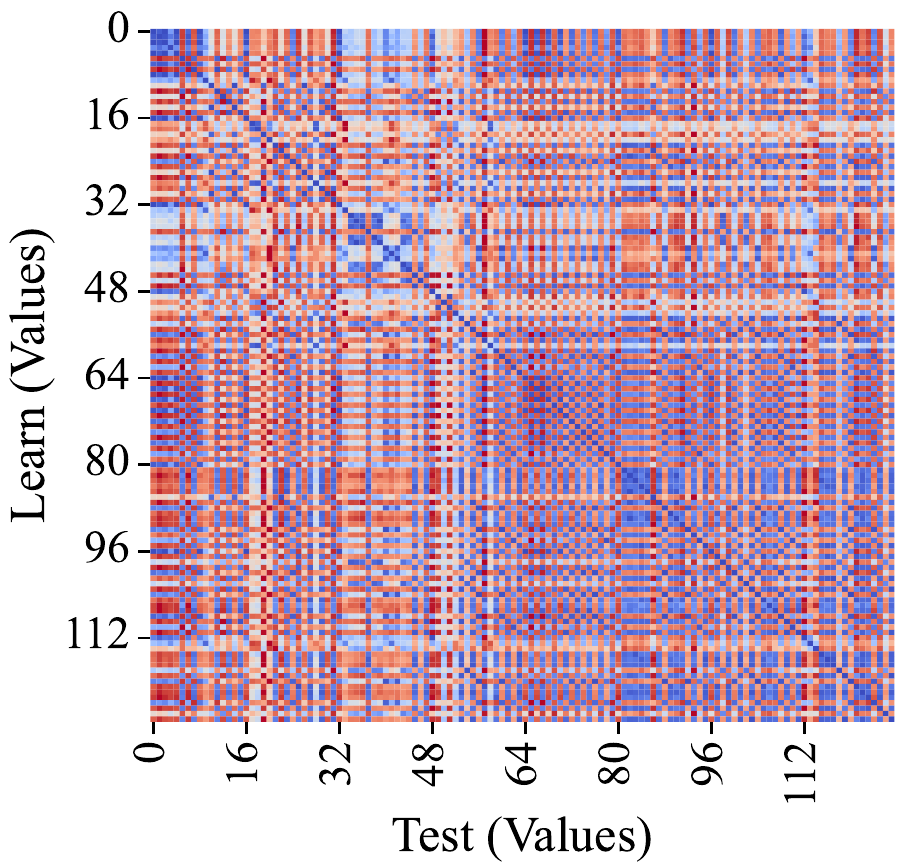}
  \caption{Optimal}
\label{fig:passive_predict_value_optimal_22.pdf}
\end{subfigure}%
\begin{subfigure}{.3\textwidth}
  \centering
  \includegraphics[width=0.95\linewidth]{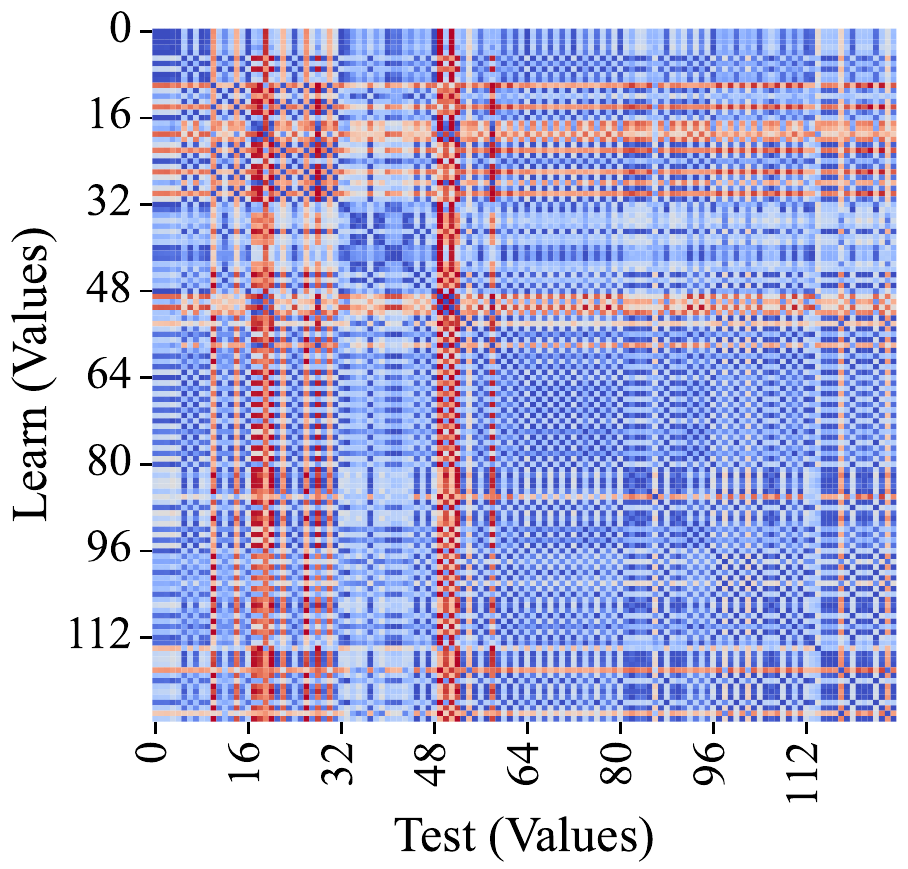}
  \caption{Default}
\label{fig:passive_attack_predict_size_10_qubit_layout}
\end{subfigure}%
\begin{subfigure}{.3\textwidth}
  \centering
  \includegraphics[width=1.12\linewidth]{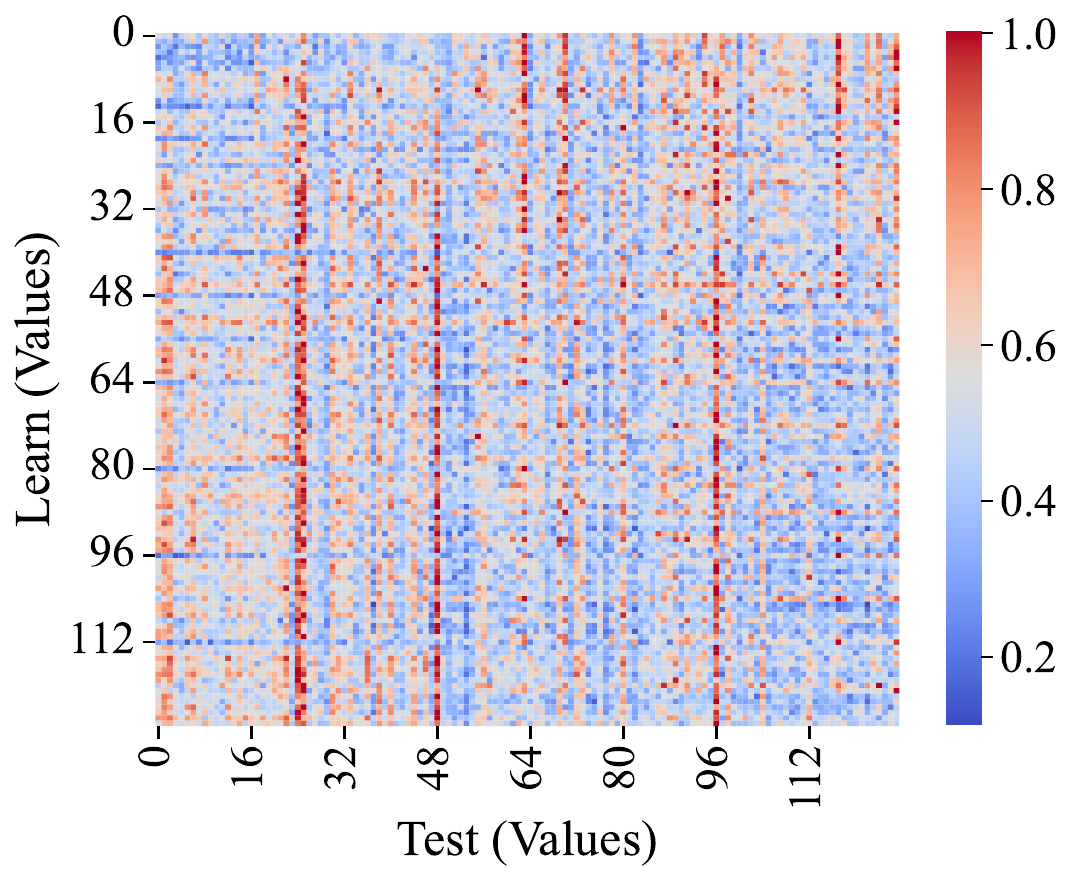}
  \caption{Non-optimal}
\label{fig:passive_predict_value_non-optimal_22}
\end{subfigure}%

\caption{MSE Matrix for Listening Circuit with $22$ Qubits.}
\label{fig:passive_predict_value_attack_qubit_selection_matrix_comparison}
\end{figure*}

As shown in Fig.~\ref{fig:passive_predict_value_attack_qubit_selection_matrix_comparison}, the differences in $Acc\_1$ are significant for a fixed number of qubits. For instance, with the Optimal Strategy using $22$ qubits, the attacker achieves $100\%$ $Acc\_1$. In contrast, the Default Strategy results in an accuracy of $86.71\%$, while the Non-optimal Strategy yields only $46.87\%$.

This experiment successfully predicts the victim’s output with high accuracy and confidence, highlighting the critical impact of qubit allocation strategies and attack circuit size on overall attack performance.

\newpage
\vspace{5pt}
\section{Conclusion} ~\label{Future}
In this paper, we focused on introducing the novel active SWAP attack that stealthily damages the victim circuit from a distant location. This attack has the varying levels of damage ranging from hard to detect minor attacks that lowers victim's output accuracy by $20-40\%$ to critical attacks that limits the capabilities of victim circuits that lowers output accuracy by $>80\%$ depending on the positions of the attack qubits relative to the victim's circuit. 
This proposed attack renders existing defense strategies ineffective such as topological distance between circuits and anti-virus detection for repetitive CNOT gates as active SWAP attack circuit is able to induce the crosstalk disruption from a distance with fewer gates.
Additionally, we also proposed an alternative Passive SWAP Attack that exploits improper qubit allocation of the victim's circuit that results in unwanted crosstalk leakage of sensitive information. 
This attack method is verified against $32$ benchmarked Simon's circuits with the ability to predict the victim's circuit with $100\%$ accuracy. We also explored the trade-off between attacker's circuit size and the prediction accuracy and confidence showcasing that a stealth circuit that is as small as $6.25\%$ the maximum size of the victim's circuit is still able to achieve $100\%$ accuracy albeit with relatively lower prediction confidence.

Beside \textit{ibm\_kyoto} and \textit{ibm\_brisbane}, future work could explore migrating the proposed attacks to other quantum devices such as IQM Garnet through AWS BraKet. In addition to the empirical observations presented in this paper, quantum researchers may be interested in further uncovering the physical principles behind this phenomenon.

\ \\
{\bf Acknowledgment:} The authors like to thank Dr. P\'{e}ter K\'{o}m\'{a}r (Applied Scientist, Amazon Braket) for his valuable inputs while preparing this paper.

\bibliographystyle{ieeetr}
\bibliography{references}

\end{document}